# Sub - THz Radiation from Dielectric Capillaries with Reflectors

K. Lekomtsev[1*], A. Aryshev[2], A.A. Tishchenko[3], M. Shevelev[2], A.A. Ponomarenko[3],
P. Karataev[1], N. Terunuma[2,4] and J. Urakawa[2]

[1]John Adams Institute at Royal Holloway University of London, Egham, Surrey, TW20 0EX UK
[2]KEK: High Energy Accelerator Research Organization, 1-1 Oho, Tsukuba, Ibaraki, 305-0801 Japan
[3]National Research Nuclear University MEPhI, Kashirskoe sh. 31, Moscow, 116409 Russia
[4]SOKENDAI: The graduate University for Advanced Studies, 1-1 Oho, Tsukuba, Ibaraki, 305-0801 Japan

**Abstract** In this report we present experimental investigations of THz radiation generated from a corrugated and a non-corrugated capillary with reflectors, using a femtosecond electron beam of LUCX accelerator at KEK, Japan. We discuss measurements of radiation angular distributions and their comparison with Particle In Cell simulations, and also investigate an off-central propagation of the beam in the capillaries based on experimental measurements and simulations.

1. Introduction

Further advances in development of linac based, tunable and narrow band intense sources of THz radiation are very important for their potential applications in the fields of spectroscopy, imaging, and security [1, 2]. Over the last several years THz radiation sources based on circular waveguides with dielectric loading or metallic corrugation have been extensively investigated theoretically [3] and experimentally [4, 5]. Selective mode excitation in a dielectric capillary using multi-bunch electron beam was tested experimentally and described in [6]. In addition to THz radiation production, circular waveguides can also serve as simple, high gradient and ultra-compact accelerating structures. Together with high repetition short electron beams they can have a transformative impact for free electron lasers as well as for linear colliders [7].

Two structures that are described in this paper are composite capillaries with a constant and a variable internal radius, on the outside surface both structures are partially covered by reflectors. The reflectors were used to increase the THz radiation yield at large angles (~ 90 deg.) with respect to the beam propagation direction. Manufacturing accuracy checks and the Particle In Cell (PIC) simulations of the radiation directivity patterns were discussed in [8]. An initial measurement of the polar angular dependence of the THz radiation from the corrugated capillary and its comparison with a PIC simulation were discussed in [9]. In this paper, we present a comprehensive analysis of both polar and azimuthal angular distributions measured for the central and off-central beam propagations inside the corrugated and the non-corrugated capillaries. We also discuss impact parameter dependencies. The aim of this paper is to demonstrate that coherent Smith Purcell Radiation (SPR), with the intensity comparable to that of coherent Transition Radiation (TR), can be generated as a result of introducing the corrugation in a capillary with a dielectric layer and can be emitted through the outer dielectric boundary not covered by the reflector.

2. Experimental setup.

THz radiation measurements were carried out at the KEK LUCX (Laser Undulator Compact X-ray source) facility [10-12]. LUCX can produce a train of up to 4 bunches, each with 90 μm (300 fs) duration, 20 pC charge and 200 μm transverse size. Short electron bunches were generated in the RF gun via photocathode illumination by femtosecond laser pulses, and further accelerated to an energy of 8 MeV in 3.6 cell RF gun. The targets were placed in the vacuum chamber which was installed right after the RF gun. The radiation was observed from the chamber through a quartz vacuum window with an effective aperture of 100 mm. Target positioning inside the chamber was done using 5-axis manipulator system, which could perform fine position adjustment in 3 orthogonal directions and over two rotational angles. Prior to radiation measurements the capillaries were aligned with respect to electron beam using forward bremsstrahlung that appeared due to a direct interaction of the electron beam with the capillaries' material. A Schottky Barrier Diode (SBD) detector with spectral sensitivity in the region 320 – 460 GHz was used for the measurements of the radiation distributions. The characteristics of the measurement setup are shown in Table 1.

A schematic view of the experiment is shown in Fig. 1, a beam consisting of a single bunch was used in the measurements. Fig. 1 (a) shows a top view of the corrugated capillary's cross-section, Fig. 1(b) shows a front view of both capillaries as they were positioned during the experiment. The schematic view does not show holders that were used to fix the capillaries on the base plate held by a vacuum manipulator, the radius of the holders' clear aperture was $r_1$ (Fig. 1(b)). In the first set of measurements the beam was propagating through the centre of either of the capillaries. In the second set of measurements it was propagating at an off-set, $h = 1.5$ mm. In both cases the SBD detector was measuring a horizontal polarization component of the radiation (along axis z in Fig. 1).

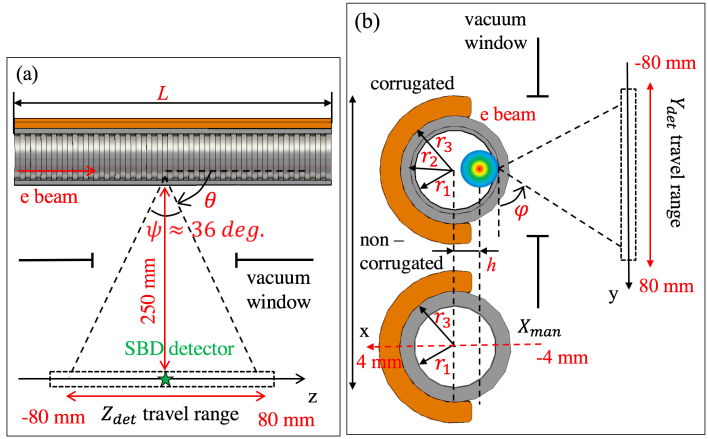

| Beam parameter | Value |
|---|---|
| Energy | 8 MeV |
| Charge | 20 pC |
| Transverse size | 200 x 200 um |
| Longitudinal size | ~ 0.09 mm |
| Capillary parameter | Value |
| Corrugation period | 1 mm |
| r1, r2, r3 | 2; 2.2; 2.7 mm |
| L | 30 mm |
| Capillary material | Fused Quartz |
| Reflector material | Copper |
| Impact parameter, $h$ | centre: 0 mm off–centre: 1.5 mm |
| Detector parameter | Value |
| Frequency range | 320 – 460 GHz |
| Wavelength range | 0.94 – 0.65 mm |
| Response time | sub-ns |
| Antenna gain | 25 dB |
| Input aperture | 4 x 4 mm |
| Video sensitivity | 1250 V/W |

Fig. 1 Schematic view of experimental geometry.       Table 1 Experimental parameters.

The positions of the Cherenkov and SPR peaks for the corrugated capillary satisfy the following dispersion relation [13]:

$$cos\theta = \frac{2\pi m}{kd} + \frac{1}{\beta\sqrt{\varepsilon(\omega)}}; \qquad (1)$$

where $\theta$ is the polar angle depicted in Fig. 1(a), $\beta$ is the bunch speed in terms of the speed of light, $k$ is the wave number in the dielectric, $d$ is the groove period, $m$ is a diffraction order and $\varepsilon(\omega)$ is the Quartz dielectric permittivity as a function of frequency. The value $m = 0$ corresponds to the Cherenkov peak, and $m = \pm n$; $n = 1,2,3$ ... correspond to the peaks of SPR. The relation (1) was originally derived for a corrugated channel in an infinite dielectric and applicable for the radiation propagation inside a material, but it may be used for the angles $\theta$ close to 90 deg. allowing for the Snell's law of refraction. In the case of the capillary with a finite outer boundary the equation (1) is only applicable for the angles $\theta = 90 \pm$ 30 deg., which do not lie in the region of full internal reflection, assuming that the average refractive index of fused quartz in THz range is $n = 1.96$ [14].

## 3. Discussion of results.

Before investigating the radiation from the capillaries, we measured the angular distribution of coherent TR emitted from a facet of a copper octagon with dimensions ($60 \ x \ 7 \ mm^2$) oriented at 45 deg. to the beam propagation direction. This surface was chosen due to unavailability of a wider flat surface that would have been preferential. One should not expect, however, a significant distortion of the radiation distribution and a large suppression of TR intensity for an electron beam with a typical field interaction region of $\gamma\lambda/2\pi \approx 2.5$mm, calculated for the parameters $\gamma = 16$ and $\lambda = 1$ mm. The polar angular distributions of the TR and the THz radiation from the corrugated capillary for the central and off-central beam propagations are shown in Fig. 2. The off-central beam propagation yields a 10-fold increase in the radiation intensity with the maximum comparable to that of TR. Fig.3 shows the distributions of the THz radiation for the off-central propagation of the beam in the corrugated and the non-corrugated capillaries. The maximum radiation intensity was normalized to unity. The fact that we see a measurable level of radiation coming from the non-corrugated capillary may be explained by its composite structure. It was constructed as a set of identical cylindrical rings, each with $\pm 50 \mu m$ manufacturing accuracy. Consequently, the capillary had irregularity in the internal radius on the same scale, which, in turn, generated SPR of a reduced intensity.

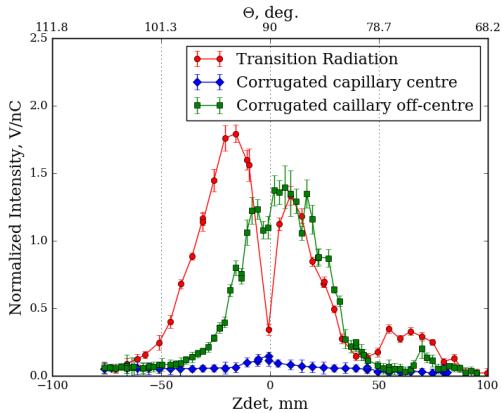
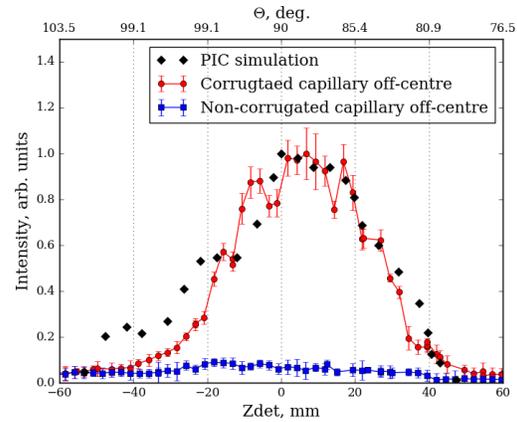

Fig. 2 Crosscheck with Transition Radiation.    Fig. 3 Radiation distribution along axis z.

The red curve in Fig. 3 was compared with the PIC simulation performed using CST Particle Studio (PS) PIC solver. The geometry used for the simulation was identical to the realistic one except for the capillary holders that were not taken into account. The beam parameters were chosen to be the same as during the time of the experiment. The simulated power distribution from the corrugation was obtained by calculating the power spectrum of the emitted THz radiation in the frequency range 245 – 365 GHz from the corrugated capillary, and the non-corrugated one with the inner and outer radii equal to $r_2$ and $r_3$, respectively (Table 1). The non-corrugated capillary considered in the simulation had the internal radius $r_2$, which was the larger value out of the two internal radii values of the corrugated capillary. It was done to simulate the radiation coming from the corrugation only and compare it with the experimental curve "Corrugated capillary off-centre" in Fig. 3. To obtain the angle $\theta$ corresponding to the direction of propagation for each frequency the radiation directivity patterns were calculated using the PIC simulation. The angular distribution was, then, recalculated into the radiation distribution along the detector travel range (axis z in Fig. 1). The power distribution of the radiation coming from the corrugation only (Fig. 3, black dots) was obtained by subtracting the simulated power distribution of the non-corrugated capillary from the simulated power distribution of the corrugated capillary as functions of $Z_{det}$. According to the dispersion relation (1) a frequency of 300 GHz can be measured only when the detector is at the position $Z_{det} = 0$, the direction towards the negative values of $Z_{det}$ corresponds to lower frequencies and the opposite direction corresponds to higher frequencies. Consequently, the measured distribution in Fig. 3 has a broad maximum that lies in the range of frequencies $300 - 320$ GHz.

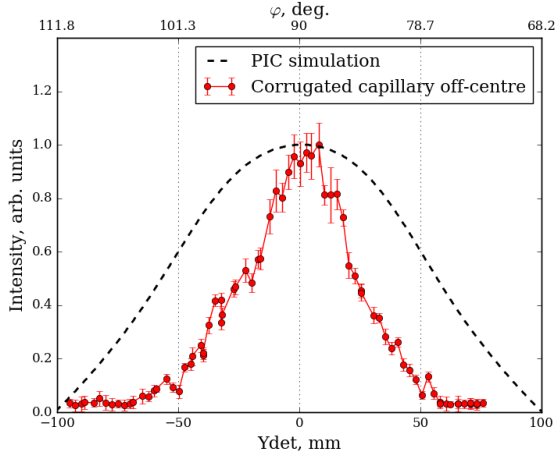

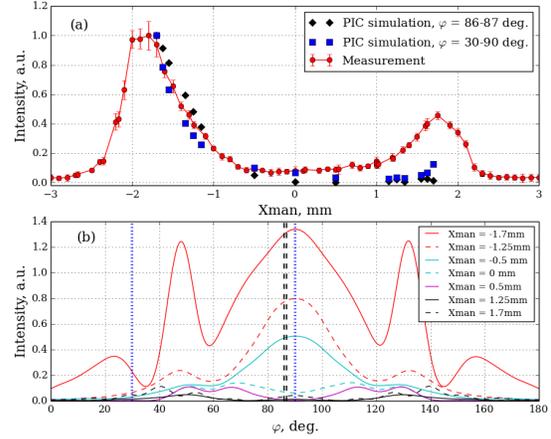

Fig. 4 Radiation distribution along axis y, $Z_{det} = 0$ and frequency is 300 GHz.

Fig. 5 Impact parameter dependence and far field patterns at 300 GHz.

Fig. 4 demonstrates an azimuthal distribution of the radiation for the off-central propagation in the corrugated capillary. The black dashed curve was calculated using the standard far-field post-processing available in CST PS. The field values at a given frequency and for the distance $d \gg \gamma^2 \lambda$ from the target are extrapolated from the field values at the border of calculation domain using the technique described in [15]. The minimum distance from the targets to the detector was 250 mm, which for $\gamma = 16$ and $\lambda = 1$mm is very close to the parameter $\gamma^2 \lambda$ that approximates a border between far and near field. For 300 GHz it was not feasible to calculate the radiation pattern exactly using a calculation domain that would cover an entire radiation propagation region due to the limit on computing resources, and for that reason the far-field monitor was used as an approximate radiation distribution pattern. The measurement in Fig. 4 demonstrates a narrower radiation distribution than for the simulation. It may be explained by a narrow angular acceptance of the SBD detector, which was approximately 1 deg. compared to 36 deg. of the scanning range along the axis z. The efficiency of radiation detection is reduced even more when the detector travels further away from the centre of the translation stage.

To study a dependence of the radiation production efficiency on the transverse position of the beam we performed an impact parameter scan by moving the corrugated capillary along the axis x (Fig. 1) using the vacuum manipulator. The scan is shown in Fig. 5 (a). We observed the maximum of the intensity when the beam was starting to propagate through the aperture, roughly at 300 $\mu$m from the corrugation. When the beam was propagating further towards the reflector, at first we observed a plateau of the radiation intensity, then the intensity increased again, but not up to the same level. This intensity inequality was observed due to the target not being symmetric. In addition, the simulations showed that the transverse shift of the beam with respect to the centre significantly changes the radiation directivity pattern (Fig. 5 (b)). The simulated impact parameter dependence was obtained by integrating a narrow region within the black dashed lines (Fig. 5 (b)), which showed the angular acceptance of the detector. During the measurements, the detector was shifted by 15 mm with respect to its central position along the axis y ($Y_{det} = 15$ mm). A good agreement between the simulation and the measurement was achieved for the negative values of x. On the contrary, for the positive values of x there was a large discrepancy, the simulated radiation intensity stayed at the minimum level, but the measurement showed an intensity increase. A wider integration region that covered most of the features of the symmetric radiation patterns, for example $\varphi = (30, 90)$ deg. (the blue dotted lines in Fig. 5(b)), described the measurement better, but the simulated maximum intensity was still smaller for the positive values of x. This discrepancy may be explained by the fact that the target holders were not considered in the PIC simulations. In addition, in the simulations a beam impact parameter was constant, however in the experiment there could have been deviations from the constant value of the impact parameter, leading to a slight change in the directivity pattern and an increased yield of radiation for the positive values of x.

## 4. Conclusions

We studied THz radiation produced by a femtosecond electron beam and the dielectric capillaries with and without corrugation. The measured radiation intensity from the corrugation was comparable to that of coherent TR. The spectral sensitivity of the detector, the simulations and the dispersion relation (1) confirmed that the measured radiation was indeed sub-THz SPR produced due to periodicity. The non-central propagation of the beam in the corrugated capillary yielded 10-fold increase in the radiation intensity. Issues that remain to be considered in the future work are further investigations of how sensitive is the radiation directivity pattern to the transverse beam position; studies of the beam dynamics during and after propagation through the capillaries; and spectral measurements for a single and a multi-bunch beam, similar to the measurements described in [16].

### Acknowledgements

Authors would like to thank S. Araki, M. Fukuda and Y. Honda for valuable help and support of the experiment. This project has received funding from the European Union's Horizon 2020 research and innovation programme under the Marie Sklodowska-Curie grant agreement No 655179; work was performed by the international collaboration AGTaX; supported by the Photon and Quantum Basic Research Coordinated Development Program from the Ministry of Education, Culture, Sport and Technology, Japan; the Leverhulme Trust through the International Network Grant (IN-2015-012). A.T. and A.P. were partly supported by the Competitiveness Programme of National Research Nuclear University "MEPhI" and by the Ministry of Science and Education of the Russian Federation, grant № 3.2621.2017/PCh.